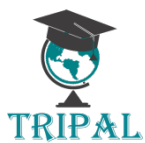

**Journal of Economics and Financial Analysis**

Type: Double Blind Peer Reviewed Scientific Journal
Printed ISSN: 2521-6627 | Online ISSN: 2521-6619
Publisher: Tripal Publishing House | DOI:10.1991/jefa.v2i2.a16

Journal homepage: www.ojs.tripaledu.com/jefa

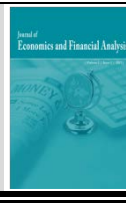

# Factors Influencing Cryptocurrency Prices: Evidence from Bitcoin, Ethereum, Dash, Litecoin, and Monero

## Yhlas SOVBETOV[*]

*Department of Economics, London School of Commerce, United Kingdom*

**Abstract**

*This paper examines factors that influence prices of most common five cryptocurrencies such Bitcoin, Ethereum, Dash, Litecoin, and Monero over 2010-2018 using weekly data. The study employs ARDL technique and documents several findings. First, cryptomarket-related factors such as market beta, trading volume, and volatility appear to be significant determinant for all five cryptocurrencies both in short- and long-run. Second, attractiveness of cryptocurrencies also matters in terms of their price determination, but only in long-run. This indicates that formation (recognition) of the attractiveness of cryptocurrencies are subjected to time factor. In other words, it travels slowly within the market. Third, SP500 index seems to have weak positive long-run impact on Bitcoin, Ethereum, and Litecoin, while its sign turns to negative losing significance in short-run, except Bitcoin that generates an estimate of -0.20 at 10% significance level.*

*Lastly, error-correction models for Bitcoin, Etherem, Dash, Litecoin, and Monero show that cointegrated series cannot drift too far apart, and converge to a long-run equilibrium at a speed of 23.68%, 12.76%, 10.20%, 22.91%, and 14.27% respectively.*

---

[*] Research Fellow, London School of Commerce, Chaucer House, White Hart Yard, London SE1 1NX, United Kingdom.
Email: ihlas.sovbetov @lsclondon.co.uk
Tel: +44 (0)207 403 1163/ Ext.364

## 1. Introduction

Cryptocurrencies have become one of the most trending topics in recent economic and financial issues. Since Dotcom crisis, the commerce on internet (e-commerce) has been rapidly increasing and retail industries have been undergoing a revolution as internet sales are booming with more and more tech-savvy consumers go online to shop. The appetite of stock market investors for e-commerce shares seemed insatiable as investments on internet retailer were massively oversized, despite fears over the future of the Internet after the dotcom bubble burst and serious concerns about the safety of online shopping by credit cards. Until birth of first cryptocurrency -Bitcoin- in 2009, the online commerce was mainly intermediated by financial institutions serving as trusted third parties to process electronic payments. Although this system was well enough for most transactions, it was working very slowly due to controls of financial institution (problem of privacy and trust) and it was somewhat cost (transaction and commission costs).

It triggered emerge of decentralized cryptocurrencies that bypass financial controllers, thus, transactions are very fast, smooth, and has zero cost. A cryptocurrency is defined as *"a digital asset designed to work as a medium of exchange using cryptography to secure the transactions and to control the creation of additional units of the currency"*. First ever use of cryptocurrency in online trade was on May $22^{nd}$ 2010 by Laszlo Hanyecz, a computer programmer from Florida, for two pizzas with the amount agreed at 10,000 Bitcoins (Yermack, 2013), which would be equivalent to $155.80 million today (December 2017).

In 2017, the popularity and use of cryptocurrencies has increased dramatically. People are *"investing"* vast sums of money into *"assets"* that have no history of producing revenue, and those assets are rising in price only because other people are also pouring money into them. Billions of dollars have been poured into more than 1,000 new digital coins issued by start-ups in 2017. These coins mimic the construction of Bitcoin, meaning they can be freely traded on digital exchanges and have no central bank standing behind them. This has raised many doubts and questions about current and future of decentralized cryptocurrencies. There are two major views. One side argues that it is a bubble with no real assets that inevitably will end with burst. The other side opines that cryptocurrency markets will become an avenue that will give millions of people an opportunity to participate in a global financial network worth tens of trillions of dollars. From young millennials in developing nations with small savings and big ambitions to mom-and-pop business owners looking to reinvest some profits in promising crypto-projects, these kinds of people will be the backbone of this industry.

This also has increased interest of cryptocurrencies in economics and financial research sphere. Although their literature is scant, number of empirical research are growing remarkably. In this respect, we also conduct a study that examines price influences of cryptocurrencies both in short- and long-run over 2010-2018 using ARDL technique on weekly basis. As statistical data of cryptocurrency are newly established, we build the *"Crypto 50"* index with its total trading volume and volatility to be used in our analysis.  This index is comprised of top 50 cryptocurrencies according to their market capitalisation rank. We sample most common five digital currencies such as Bitcoin, Ethereum, Litecoin, Dash, and Monero, and we scrutinize how these currencies interaction with stock market (SP500 index), gold prices, and with macroeconomic indicators (Interest rate) both in short- and long-run.

## 2. Characteristics of Cryptocurrency

A cryptocurrency is a digital or virtual currency that uses cryptography for security. A cryptocurrency is difficult to counterfeit because of this security feature. A defining feature of a cryptocurrency, and arguably its most endearing allure, is its organic nature; it is not issued by any central authority, rendering it theoretically immune to government interference or manipulation. It is designed from the ground up to take advantage of the internet and how it works. Instead of relying on traditional financial institutions that verify and guarantee your transactions, cryptocurrency transactions are verified by the user's computers logged into the currency's network. Since the currency is protected and encrypted, it becomes impossible to increase the money supply over a predefined algorithmic rate.

One cryptocurrency, in particular, has entered the public lexicon as the go-to digital asset: Bitcoin, often is regarded as father of cryptocurrencies and all other cryptocurrencies are referred as altcoins. Since 2009, the finance world has been watching the crackerjack rise of Bitcoin with a combination of fascination and, in many cases, severe skepticism. Characteristics of Bitcoin make it fundamentally different from a fiat currency, which is backed by the full faith and credit of its government. Fiat currency issuance is a highly centralized activity supervised by a nation's central bank. On the other hand, the value of a Bitcoin is wholly dependent on what investors are willing to pay for it at a point in time. It uses peer-to-peer blockchain network (chronologically arranged chain of blocks where each block has a list of transactions information) where all members are equal and there is no central server that tells everyone what to do (Nakamoto, 2008). This decentalisation is maintained on Satoshi Nakamoto's (2008) idea of combining *"proof of work"* (PoW) with other cryptographic techniques. The PoW, mathematically, is a hash function with a large number of answer variances, the so-called *"beautiful"* hash is considered to be the one that is characterized by starting with 15 zeros. The hash of each block is algorithmically directly linked to the previous block. That is, if we hypothetically represent the hash function in the form,

$$Hash\ of\ current\ block = f(\theta, \phi, Z)$$

where θ is the hash of the previous block; $\phi$ is current difficulty level; and *Z* is a random key uniquely specific to the current block. This indicates that each subsequent block is inextricably linked to the previous one due to θ, and if any dishonest miner at some point decides to generate an invalid block, the other network members will not confirm it, because the hash of the previous block will not be used in it. And if spammer decides to change the hash of the previous block, then he will have to do this for the previous one as well, and so on until the genesis block (the very first block created by Satoshi Nakamoto himself). It would be incredibly time consuming to comb through the entire ledger to make sure that the person mining the most recent batch of transactions hasn't tried anything funny. This will require huge amount of work, which at the moment is almost beyond the power of one person or even a large organization. Therefore, PoW also maintains defense mechanisms for cryptcurrencies against hacking.

However, PoW miners invest into advanced computer machines that 24/7 works (consuming energy) with the goal of validating transactions (solving hashes) and creating new blocks. Once it finds *"beautiful"* it declares that the block is resolved and every miner gets reward (bitcoins) proportional to their work spent on solving the hash. Therefore, cryptocurrency mining under PoW protocol is painstaking, expensive, and only sporadically rewarding. Alternatively, many altcoins started to use *"proof-of-stake"* (PoS) protocol which is more cost effective (cheaper) and eco-friendly (greener) comparing to PoW that requires a lots of computer energy consumption to solve mathematical algorithmic hashes. In case of PoS, miners do not need expensive computer machines, the creator of a new block is chosen in a deterministic way, depending on its wealth, also defined as stake.

Majority of cryptocurrencies has roof limit of production. It means that supply of cryptcurrencies would decrease over time and under *ceteris paribus* condition should lead to higher price (inflation). However, unlike centralized fiat currencies, the cryptocurrencies are unique since their block reward schedule is public. It means that public already knows the approximate date of each decrease (or reward halving). Thus all expectation should have been purchased by the market, and therefore shrunk in supply should not affect cryptocurrencies trading price. For instance, Bitcoin's first block halving happened on 28[th] November 2012. The block reward dropped from 50 BTC per block to 25 BTC per block. The price later climbed to $260 BTC in April 2013, followed by $1,163 BTC in November 2013. It is unclear, however, whether these price rises were directly related to the block reward halving. In this research, we investigate factors that influence cryptocurrency prices both in short- and long-run.

## 3. Literature Review

The cryptocurrency market has seen an unprecedented level of interest from investors in 2016. Bitcoin, the world's largest digital currency, has risen more than 1,500 percent since the start of 2017. However, the market is significantly more complex than the public lexicon might suggest. And while there have been plenty of studies examining the future of Bitcoin and its volatility (Polasik et al. 2015; Letra, 2016; Bouoiyour and Selmi, 2016; Katsiampa, 2017; Chiu and Koeppl, 2017; Chu et al. 2017), there have been few that explore the broader cryptocurrency market and how it is evolving. Bitcoin is currently trading at around $16,000; at the beginning of the year, Bitcoin price was at $1,000, raising warnings from some analysts and prominent financial figures that it's a bubble. The currency is extraordinarily volatile despite its recent ever-peaking performance, rising by thousands of dollars in value on one day only to fall by even more the next. Katsiampa (2017) estimates the volatility of Bitcoin through a comparison of GARCH models and finds that the AR-CGARCH model gives the most optimal fit. He underlines that the market is high speculative. Bouoiyour and Selmi (2016) study daily Bitcoin prices using an optimal-GARCH model and show that the volatility has decreasing trend comparing pre- and post-2015 data. Even tough, they still observe significant asymmetries in the Bitcoin market where the prices are driven more by negative than positive shocks. Likewise, Dyhrberg (2016) investigates the asymmetric GARCH methodology to explore the hedging capabilities of Bitcoin and he finds that it can be used as a hedging tool against stocks in the Financial Times Stock Exchange Index and against the American dollar in the short term.

On the other hand, El Bahrawy and Alessandretti (2017) examine behaviour of entire market (1469 cryptocurrencies) between April 2013 and May 2017. They find that cryptocurrencies appear and disappear continuously and their market capitalization is increasing (super-)exponentially, several statistical properties of the market have been stable for years. Particularly, market share distribution and the turnover of crytocurrencies remain quite stable.

There is a wide agreement on that the cryptocurrencies will not only affect the trading practices of different countries and business organizations, but they will also affect the dynamics of international relations. There are still a lot of people who are never accommodating the idea that cryptocurrencies will revolutionize how we do businesses. They can't figure out how the whole blockchain technology and other annexes work. Plus, advancements in technology are introducing digital tools that companies can use to better interact with their customers. A rising shift from traditional platforms to digital platforms has also brought about an abundant supply in data from sources like social media, mobile devices, online retail platforms, etc. Due to technology advancements in the areas of gathering, storing, and sharing data, large sets of data are easily shared among companies in every sector and country for little to no costs. The widespread accessibility of data has also brought about concerns over data privacy of individuals and their online transactions. Because every transaction or activity carried out online leaves a digital trail, individuals are opting for more anonymous ways to use the internet and conduct online transactions. The Bitcoin cryptocurrency was introduced to address the issue of privacy concern.

Although cryptocurrencies' decentralization, anonymity of transaction, and irreversibility of payments offer plenty advantages, Brill and Keene (2014) opine that these features also attract illegal activities (cybercriminals) such as money laundering, drug peddling, smuggling and weapons procurement. This issue has attracted the attention of powerful regulatory and other government agencies such as the Financial Crimes Enforcement Network (FinCEN), the SEC, and even the FBI and Department of Homeland Security (DHS). In March 2013, FinCEN issued rules that defined virtual currency exchanges and administrators as money service businesses, bringing them within the ambit of government regulation. In May that year, the DHS froze an account of Mt. Gox – the largest Bitcoin exchange – that was held at Wells Fargo, alleging that it broke anti-money laundering laws. And in August, New York's Department of Financial Services issued subpoenas to 22 emerging payment companies, many of which handled Bitcoin, asking about their measures to prevent money laundering and ensure consumer protection. Plus, economist Kenneth Rogoff writes that Bitcoin will never supplant government-issued money because that *"would make it extremely difficult to collect taxes or counter criminal activity."* (see Bitcoin legality in Appendix table 1A).

To summarize, Poyser (2017) points three types of crypto price drivers organized into internal and external factors. Supply and demand of cryptcurrency is main internal factors that have direct impact on its market price. On the other hand, attractiveness (popularity), legalization (adoption), and few macro-finance factors (interest rate, stock markets, gold prices) can be regarded as external drivers (see figure 1).

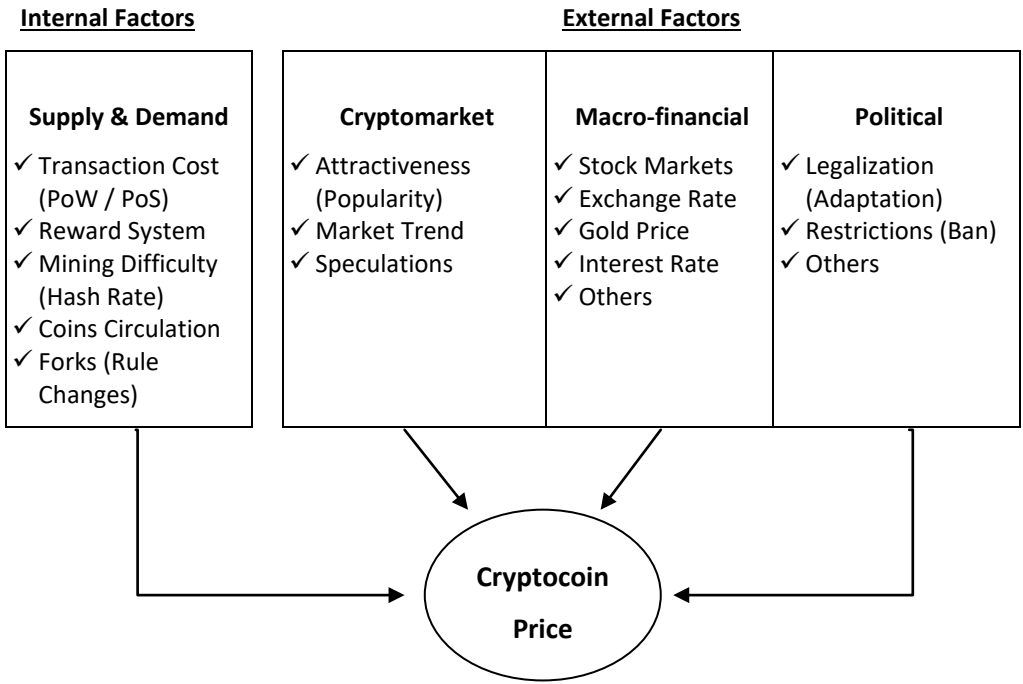

**Figure 1.** Factors that Influence Cryptocurrency Prices

In this respect, we examine short- and long-run factors that influence prices of cryptocurrencies over 2010-2018 using ARDL technique on weekly data basis. First, we build Crypto 50 index by sampling top 50 cryptocoins that have proportional contribution to market capitalization weights. Thus, we derive few cryptomarket factors such as total market capitalization, trading volume, and volatility. We use these factors as explanatory variables for cryptocoin price movements alongside with attractiveness and control variables such as stock market movements, gold prices, and interest rates. In this study, we provide evidence for significant long-run role of attractiveness of cryptocurrencies in determination of their prices. We also observe a weak form of negative impact running from stock markets (SP 500 index) to cryptocurrency market, in particular Bitcoin.

The contents of the paper are organized as follows. Next section describes the data with descriptive analysis and explains methodological set up of examination. Then, we present our key findings including our comments and suggestions. The final section gives concluding remarks of the study.

## 4. Data and Methodology

The literature about economics of cryptocurrency is scant as the topic just recently gained focus on research fields. We contribute to this context by examining factors that influence prices of most common five cryptocurrencies over 2010-2018 with weekly data. For this examination, we define our econometric set up as following.

$$P_{c,t} = \beta_0 + \sum_{i=1}^{m} \gamma_i P_{c,t-i} + \sum_{i=1}^{3} \beta_i \Omega_t + \beta_4 ATR_{c,t} + \sum_{i=1}^{k} \alpha_i Z_{i,t} + \varepsilon_t \qquad (1)$$

where $m$ is optimal lag length which is determined by information criteria; $P_{c,t}$ is endogenous variable in the system and it denotes price of cryptocurrency "c" in natural logarithmic form at month $t$. We treat all other variables in the system as exogenous variables. The $\Omega$ represents three cryptomarket variables of $MARP_t$, $MARV_t$, and $MARS_t$ that are Crypto 50 index price (see section 4.1), its trading volume, and its volatility at week $t$; and $ATR_{c,t}$ is attractiveness of currency "c". Plus, we also account $k$ set of control variables of $Z_i$ such as stock market (proxied by SP500 index), exchange rates (EURO/USD), the U.S. interest rates, and world gold price.

Data for cryptocurrencies are gathered from BitInfoCharts[1] website; price of SP500 index is retrieved from Yahoo Finance[2], and macroeconomic data are obtained from World Bank[3]. The attractiveness of cryptocurrency is proxied by its Google search frequency; we derive related data from Google search trends[4].

### 4.1. Building Crypto 50 index

First of all, we sample big 50 market capped cryptocurrencies (these 50 cryptocoins forms about 92% of entire cryptomarket). We derive data for market capitalization, trading volume, opening-closing prices, and high-low prices from Coinsmarketcap[5]. Then, we calculate weight of each cryptocoins in the index on the basis of their market capitalization. We establish Crypto 50 index (*CRX50*) price by summing all fifty weighted-prices as following methodology.

---

[1] BitInfoCharts - https://bitinfocharts.com/

[2] Historical Prices - https://finance.yahoo.com/world-indices

[3] https://data.worldbank.org/indicator/

[4] https://trends.google.com/trends

$$CRX50\ INDEX\ Price_t = MARP_t = \sum_{i=1}^{50} \frac{MC_{i,t}}{MC_{CRX50,t}} P_{i,t} \qquad (2)$$

where $MC_{i,t}$ and $P_{i,t}$ are market capitalization and price of cryptocoin *i* at time *t* respectively; and $MC_{CRX50,t}$ is total market capitalization of cryptocoins, that forms *CRX50*, at time *t*. We also derive daily trading volume of *CRX50* index by simply summing up trading volumes of all its constituents.

$$CRX50\ INDEX\ Volume_t = MARV_t = \sum_{i=1}^{50} VOL_{i,t} \qquad (3)$$

where $VOL_{i,t}$ is total trading volume of cryptocoin *i* at time *t*. Then, we derive daily volatility of our *CRX50* index using formula below.

$$CRX50\ INDEX\ Volatility_t = MARS_T = ln\left(\frac{P_{h,t}}{P_{l,t}}\right) \qquad (4)$$

where $P_{h,t}$ is the highest price of *CRX50* index recorder at day *t*, while $P_{l,t}$ is the lowest price of *CRX50* index recorded at day *t*. The high-low price of *CRX50* index is derived by methodology explained in equation 2.

### 4.2. Brief Overview of Cryptocurrency Market

We briefly summarize economics of cryptocurrency market by outlining key statistics. Coin Dance[6] regularly announces up-to-date and historical report statistics of cryptocurrency markets. According their most recent report, 34.4% of total market share belongs to Bitcoin, while 19.23%, 10.74%, and 1.97% shares are attributable to Ethereum, Ripple, and Litecoin respectively. Moreover, their report shows that 96.57% of cryptocurrency market involvers are males, while only 3.43% are females. The age distribution refers to ability of the cryptocurrency market to attract wide range of people from very young to very old. The report gives statistics for only 18+ ages where 8.36% of the market involvers are aged 18-24, while 45.71% and 30.62% are attributable to people aged 25-34 and 35-44 respectively. Interestingly, the share of elderly people (45+) is about 16% which provides evidence for that cryptocurrency market attracts from youngest to elderly people into financial activities. The cryptocurrency interest and affinity statistics also show that people of cryptocurrency community are mainly engaged with financial activities, pursing investment opportunities.

---

[5] https://coinmarketcap.com/

[6] https://coin.dance

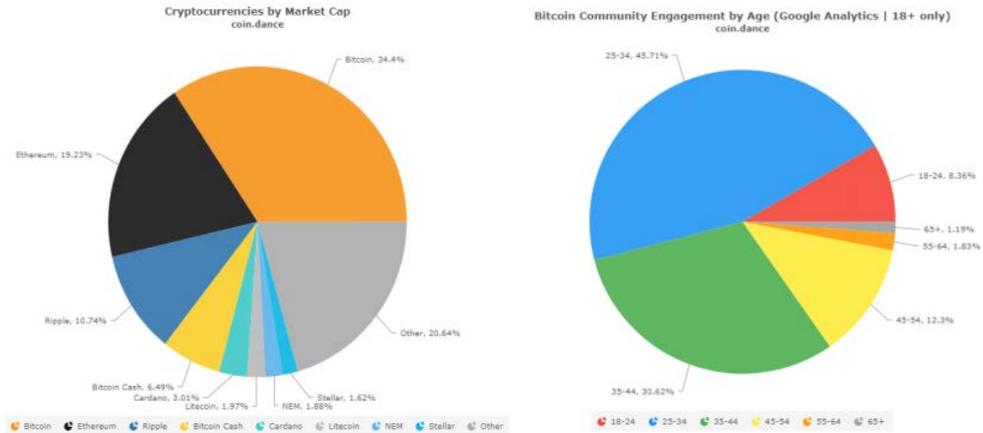

**Figure 2.** Cryptocurrency Market Share and Engagement Demographics

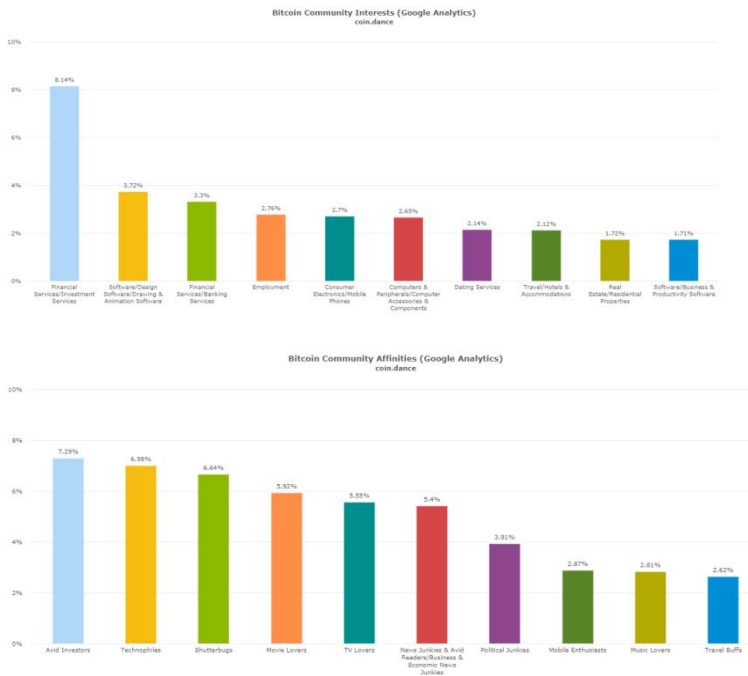

**Figure 3.** Bitcoin Community Interest and Affinity

On the other hand, Google search frequency for *"Bitcoin"* (or *"BTC"*) and *"Blockchain"* terms also shows fairly significant correlation with Bitcoin and Altcoins prices respectively (see figure 4). This seems to be a significant explanatory factor of cryptocurrency prices; therefore, we use this indicator to proxy attractiveness of cryptocurrency in this study.

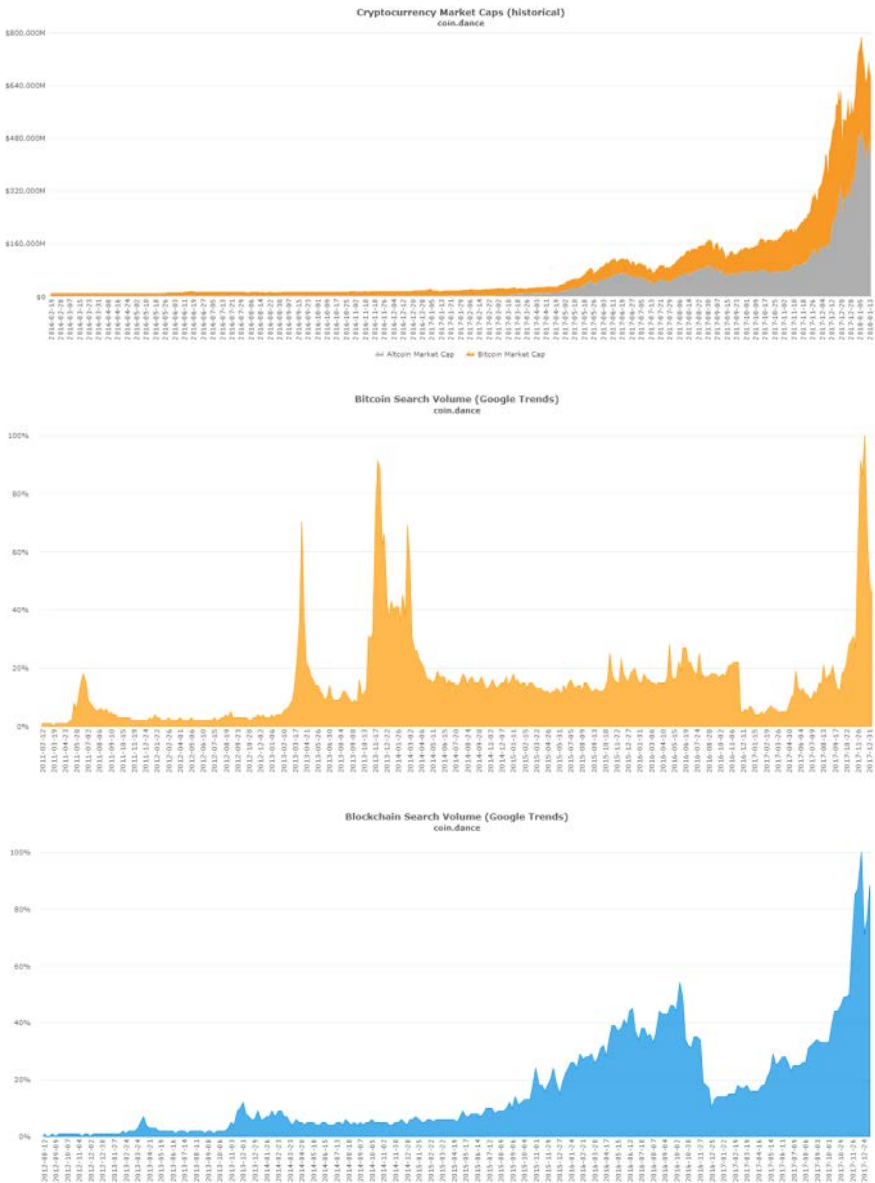

**Figure 4.** Bitcoin-Altcoin Market Cap vs Bitcoin-Blockchain Google Search Trends

To summarize, we briefly outline some cryptocurrency-specific figures in table 1, plus, we describe our data with brief abbreviations and statistics in table 2.

**Table 1.** Overview of Cryptocurrency Market

| | Bitcoin | Ethereum | Dash | Litecoin | Monero |
|---|---|---|---|---|---|
| Max. Supply | 21 million BTC | No Limit | 18.9 million DASH | 84 million LTC | No Limit |
| Total | 16.7 million BTC | 96.8 million ETH | 7.8 million DASH | 54.7 million LTC | 15.6 million XMR |
| Price (USD) | $ 14,729.86 | $ 1,082.47 | $ 1,067.01 | $ 248.93 | $ 389.18 |
| Market Cap. (USD) | $248 billion | $106 billion | $8 billion | $13 billion | $6 billion |
| Transactions / hour | 13,609 | 49,900 | 614 | 6,036 | 220 |
| Sent / hour | 121,019 BTC ($1.8 billion USD) | 558,178 ETH ($612 million USD) | 9,276 DASH ($10 million USD) | 559,875 LTC ($140 million USD) | 683,751 XRP ($1.63 million USD) |
| Avg. Transaction Value | 8.89 BTC ($131,519 USD) | 11.19 ETH ($12,273 USD) | 15.10 DASH ($16,146 USD) | 92.76 LTC ($23,174 USD) | 75.28 XMR ($29,297 USD) |
| Median Transaction Value | 0.366 BTC ($5,416.88 USD) | 0.197 ETH ($216.3 USD) | 0.605 DASH ($647.12 USD) | 10.83 LTC ($2,706.15 USD) | 9.35 XMR ($3638.33 USD) |
| Block Time | 9m 17s | 15.8s | 2m 37s | 2m 33s | 2m 0s |
| Blocks Count | 503,189 | 4,875,208 | 800,814 | 1,346,832 | 1,487116 |
| Blocks last 24h | 154 | 5460 | 548 | 562 | 713 |
| Blocks / hour | 6 | 228 | 23 | 23 | 30 |
| Reward Per Block | 12.50 BTC ($246,521 USD) | 3 ETH ($4,744 USD) | 3.60 DASH ($3,867USD) | 25 LTC ($6,321 USD) | 5.43 XMR ($2,186 USD) |
| Difficulty | $1.93114*10^{12}$ | $2.003*10^{15}$ | $70.58*10^6$ | $3.7*10^6$ | $75.8 * 10^9$ |
| Hashrate (Hash/second) | $15.58 * 10^{18}$ | $169.10 *10^{12}$ | $1.93 *10^{15}$ | $99.98 *10^{12}$ | $6.26 * 10^8$ |
| Mining Profitability/Day | 2.4364 USD | 0.1531 USD | 0.5493 USD | 0.0355 USD | 2.4905 USD |
| Wealth Distribution Top 10 addesses Top 100 addesses Top 1,000 addesses Top 10,000 addesses | 10 - 5.25% 100 - 17.89% 1000 - 34.25% 10000 - 55.66% | 10 - 10.82% 100 - 33.90% 1000 - 53.75% 10000 - 69.61% | 10 - 6.32% 100 - 15.64% 1000 - 28.53% 10000 - 92.37% | 10 - 14.44% 100 - 48.61% 1000 - 65.91% 10000 - 79.96% | 10 - 18.03% 100 - 51.17% 1000 - 71.85% 10000 - 84.29% |
| 100 Largest Transactions in Last 24h | 713,840 BTC ($10.56 billion) 24.58% Total | 1,055,897 ETH ($1.1 billion) 7.88% Total | 128,562 DASH ($137 million) 57.75% Total | 2,806,164 LTC ($701 million) 20.88% Total | 2,042,328 XMR ($794 million) 13.25% Total |
| First Block (Genesis) | 2009-01-09 | 2015-07-30 | 2014-01-19 | 2011-10-08 | 2014-04-18 |
| Genesis Info | Contained 1 transaction with 50 BTC reward by Satoshi Nakamoto, including message "The Times 03/Jan/2009 Chancellor on brink of second bailout for banks". | Contained 8893 transactions with no reward by Ethereum Foundation (team led by Vitalik Buterin) | Contained 1 transaction with 500 DASH reward by Evan Duffield. | Contained 1 transaction with 50 LTC reward by Charlie Lee | Contained 1 transaction with 17.592 XMR reward by The Monero Core Team (forked from Bytecoin) |
| Blockchain Size | 178.49 GB | 293.11 GB | 5.04 GB | 12.69 GB | 39.22 GB |
| Consensus | Proof of Work (SHA-256) | Proof of Work / Proof of Stake | Proof of Work (X11) + Masternodes | Proof of Work (Scrypt) | Proof of Work (CryptoNight) |

**Table 2.** Descriptive Statistics

| | Series | Abbr | Mean | Median | Max. | Min. | Std. Dev. | Skew-ness | Kurtosis | N |
|---|---|---|---|---|---|---|---|---|---|---|
| **PRICE** | *Bitcoin price* | BITPP | 4.43 | 5.48 | 9.72 | -2.81 | 2.86 | -0.76 | 2.86 | 390 |
| | *Ethereum price* | ETHP | 2.91 | 2.48 | 6.87 | -0.80 | 2.08 | 0.15 | 2.03 | 126 |
| | *Dash price* | DASP | 2.28 | 1.85 | 7.35 | -0.53 | 1.84 | 1.01 | 3.01 | 203 |
| | *Litecoin price* | LITP | 1.83 | 1.36 | 5.75 | 0.15 | 1.16 | 1.24 | 3.85 | 245 |
| | *Monero price* | MONP | 0.63 | -0.15 | 6.15 | -3.99 | 2.41 | 0.43 | 2.15 | 193 |
| | *EURO/USD price* | EURP | 1.25 | 1.28 | 1.48 | 1.04 | 0.12 | -0.13 | 1.65 | 390 |
| | *Gold price* | GOLP | 7.21 | 7.17 | 7.51 | 6.96 | 0.14 | 0.57 | 2.27 | 390 |
| | *SP500 price* | SPP | 7.46 | 7.53 | 7.91 | 6.96 | 0.24 | -0.23 | 1.81 | 390 |
| **MARKET CAP** | *Bitcoin mar.cap* | BITM | 12.34 | 3.47 | 297.53 | 0.00 | 34.99 | 5.67 | 39.72 | 390 |
| | *Ethereum mar.cap* | ETHM | 9.80 | 0.99 | 85.70 | 0.03 | 17.02 | 2.28 | 8.65 | 126 |
| | *Dash mar.cap* | DASM | 0.54 | 0.03 | 9.55 | 0.00 | 1.46 | 4.31 | 23.70 | 203 |
| | *Litecoin mar.cap* | LITM | 0.78 | 0.18 | 19.01 | 0.04 | 2.23 | 6.03 | 43.28 | 245 |
| | *Monero mar.cap* | MONM | 2.19 | 0.24 | 95.46 | 0.03 | 7.74 | 8.74 | 97.55 | 193 |
| **TRADING VOLUME** | *Bitcoin volume* | BITV | 4.50 | 4.04 | 10.01 | 2.25 | 1.75 | 1.15 | 3.60 | 210 |
| | *Ethereum volume* | ETHV | 3.45 | 3.03 | 8.54 | -1.61 | 2.60 | 0.07 | 2.18 | 126 |
| | *Dash volume* | DASV | -0.42 | -1.25 | 6.71 | -4.65 | 2.62 | 0.85 | 2.59 | 203 |
| | *Litecoin volume* | LITV | 2.00 | 1.26 | 8.13 | -0.37 | 1.96 | 1.31 | 3.63 | 210 |
| | *Monero volume* | MONV | 0.63 | -0.15 | 6.15 | -4.00 | 2.41 | 0.43 | 2.15 | 193 |
| **MARKET** | *CRX50 price* | MARP | 4.31 | 5.37 | 9.23 | -2.81 | 2.75 | -0.86 | 2.96 | 390 |
| | *CRX50 mar.cap* | MARM | -0.39 | 1.37 | 6.45 | -11.33 | 4.10 | -1.01 | 3.34 | 390 |
| | *CRX50 volume* | MARV | -2.04 | -2.54 | 3.99 | -4.32 | 1.97 | 1.20 | 3.51 | 210 |
| | *CRX50 volatility* | MARS | -3.11 | -3.21 | 0.04 | -9.87 | 1.07 | -0.27 | 6.74 | 390 |
| **ATTRACTIVENESS** | *Bitcoin trend* | BITA | 0.78 | 0.78 | 2.00 | 0.30 | 0.34 | 1.32 | 4.92 | 390 |
| | *Ethereum trend* | ETHA | 0.14 | 0.00 | 2.00 | 0.00 | 0.40 | 3.23 | 9.85 | 126 |
| | *Dash trend* | DASA | 0.15 | 0.00 | 2.00 | 0.00 | 0.30 | 3.42 | 10.87 | 203 |
| | *Litecoin trend* | LITA | 0.10 | 0.00 | 2.00 | 0.00 | 0.32 | 3.49 | 12.87 | 245 |
| | *Monero trend* | MONA | 0.31 | 0.30 | 2.00 | 0.00 | 0.21 | 4.75 | 33.38 | 193 |

### 4.3. Model Specification

Prior to cointegration analysis, we should make sure that variables are integrated at same degree. We examine characteristics of all series by employing Augmented Dickey-Fuller (ADF) unit root test as following.

$$\Delta \Omega_t = \theta_0 + \theta_1 T + \rho \Omega_{t-1} + \sum_{i=1}^{k} \theta_{i+1} \Delta \Omega_{t-i} + \varepsilon_t \qquad (2)$$

where $\Delta \Omega_t$ is the first difference of a variable $\Omega$; $T$ is a trend, and $\theta_1$ is its multiplier; $k$ is an optimal lag length; and $\varepsilon_t$ is a White Noise residual term. Here, ADF hypothesizes $H_0$ ($\rho$=0) against alternative ($\rho$<0), and rejection of the null confirms stationarity of $\Omega$.

We display results of ADF test in table 3 where we find majority of series are non-stationary at level. But they can be converted to a stationary through first differencing methodology. Thus, we conclude that all series are *I(1)* variables, except volatility (sigma) and few attractiveness series that are seem to be *I(0)*. Finding series are not integrated in same degree, we decide to use Autoregressive Distributed Lag (ARDL) cointegration framework which is also known as Bound testing approach. This technique is applicable for series with mixture of *I(0)* and *I(1)* variables, but none of them should be *I(2)*.

**Table 3.** Output of ADF Analysis

| | Series | Level | | | | First Difference | | | |
|---|---|---|---|---|---|---|---|---|---|
| | | Prob. | Lag | Max Lag | N | Prob. | Lag | Max Lag | N |
| PRICE | BITPP | 0.3360 | 1 | 16 | 388 | 0.0000 | 0 | 16 | 388 |
| | ETHP | 0.5106 | 2 | 12 | 123 | 0.0000 | 1 | 12 | 123 |
| | DASP | 0.9654 | 0 | 14 | 202 | 0.0000 | 0 | 14 | 201 |
| | LITP | 0.9873 | 0 | 15 | 244 | 0.0000 | 0 | 15 | 243 |
| | MONP | 0.2991 | 2 | 14 | 190 | 0.0000 | 1 | 14 | 190 |
| | EURP | 0.4698 | 0 | 16 | 389 | 0.0000 | 0 | 16 | 388 |
| | GOLP | 0.2050 | 0 | 16 | 389 | 0.0000 | 0 | 16 | 388 |
| | SPP | 0.2148 | 1 | 16 | 388 | 0.0000 | 0 | 16 | 388 |
| MARKET CAP | BITM | 0.1352 | 3 | 16 | 386 | 0.0238 | 2 | 16 | 386 |
| | ETHM | 0.1927 | 3 | 12 | 122 | 0.0003 | 1 | 12 | 123 |
| | DASM | 0.9847 | 0 | 14 | 202 | 0.0375 | 1 | 14 | 200 |
| | LITM | 0.9272 | 2 | 15 | 242 | 0.0458 | 2 | 15 | 241 |
| | MONM | 0.8451 | 0 | 14 | 190 | 0.0174 | 0 | 14 | 189 |
| TRADING VOLUME | BITV | 0.1214 | 1 | 14 | 208 | 0.0000 | 3 | 14 | 205 |
| | ETHV | 0.0231 | 0 | 12 | 125 | 0.0000 | 0 | 12 | 124 |
| | DASV | 0.6298 | 2 | 14 | 200 | 0.0000 | 1 | 14 | 200 |
| | LITV | 0.6355 | 2 | 14 | 207 | 0.0000 | 1 | 14 | 207 |
| | MONV | 0.4942 | 0 | 14 | 190 | 0.0000 | 1 | 14 | 188 |
| MARKET | MARP | 0.2931 | 1 | 16 | 388 | 0.0000 | 0 | 16 | 388 |
| | MARM | 0.1334 | 3 | 16 | 386 | 0.0000 | 2 | 16 | 386 |
| | MARV | 0.7168 | 2 | 14 | 207 | 0.0000 | 3 | 14 | 205 |
| | MARS | 0.0000 | 1 | 16 | 388 | 0.0000 | 5 | 16 | 383 |
| ATTRACTIVENESS | BITA | 0.0732 | 2 | 16 | 388 | 0.0051 | 1 | 16 | 388 |
| | ETHA | 0.0978 | 1 | 12 | 124 | 0.0308 | 0 | 12 | 124 |
| | DASA | 0.1433 | 2 | 14 | 200 | 0.0165 | 0 | 14 | 201 |
| | LITA | 0.1340 | 2 | 15 | 242 | 0.0334 | 1 | 15 | 242 |
| | MONA | 0.1143 | 0 | 15 | 190 | 0.0291 | 0 | 15 | 189 |

**Notes:** The numbers are F-statistics derived with ADF unit root test using levels and first differences. The lag length criterion is set as Schwarz (1978) Information Criterion (SIC) with automatic maximum 25 lags. The tested model includes individual effects and individual linear trends.

The bound testing methodology, is pioneered by Pesaran et al (2001), tests potential cointegration of *I(0)* and *I(1)* variables in long-run. The technique also provides some evidence for series short-run and error-correction dynamics. Thus, we recall equation 1, and adjust it in accordance to the ARDL approach following Pesaran et al (2001) as below.

$$\Delta P_{c,t} = \beta_0 + \sum_{i=1}^{m} \gamma_i \Delta P_{c,t-i} + \sum_{j=1}^{4}\sum_{i=1}^{n} \beta_{ij}\Delta X_{j,t-i} + \alpha_{ij}\Delta Z_{j,t-i} + \varphi_1 P_{c,t-1} + \sum_{j=1}^{4} \varphi_{2j} X_{j,t-1} + \varphi_{3j} Z_{j,t-1} + \mu_t \quad (3)$$

where *X* stands for four cryptocurrency-related variables of *MARP*, *MARV*, *MARS*, and *ATR*; and *Z* stands for four control variables of *SPP*, *EURP*, *GOLP*, and *INT*. The lag of dependent variable starts from 1 to its optimal lag length ($m$). However, the independent variables begin from lag zero and continue up to their optimal, i.e. $n_1$-$n_8$, which are determined by Schwartz (1978) Information Criterion (SIC).

Thus, the null hypothesis of $\varphi_1 = \varphi_{2j} = \varphi_{3j} = 0$ is tested with Wald analysis where rejection of $H_0$, under Pesaran et al (2001) lower and upper bound critical values, indicates existence of long-run cointegration between series only if the residual of equation 1.0 model ($\varepsilon_t$) is stationary. In case of justification of these requirements the Restricted Error Correction Model (RECM) can be formulated as below.

$$\Delta P_{c,t} = \beta_0 + \sum_{i=1}^{m} \gamma_i \Delta P_{c,t-i} + \sum_{j=1}^{4}\sum_{i=1}^{n} \beta_{ij}\Delta X_{j,t-i} + \alpha_{ij}\Delta Z_{j,t-i} + \lambda_c ECT_{c,t-1} + \omega_t \quad (4)$$

where $ECT_c$ is White-noise stationary residual of long-run equation 2 ($\varepsilon_t$), and $\lambda_c$ is its multiplier that is expected to be statistically significant in the range of -1 and 0 for robustness of RECM model (3). In case, $\lambda_c$ is estimated positive, then the model is suffering of serially correlated residual terms (autocorrelation problem). And if $\lambda_c$ is estimated negative but greater than 1 (in absolute terms), then the model is instable comprising of structural breaks that are needed to be controlled (Sovbetov and Saka, 2018).

## 5. Findings

Firstly, we ensure that residual of equation 1 model ($\varepsilon_t$) is stationary at 1% significance level. Next, we employ an ARDL optimal lag selection test in Eviews 9.0 software for equation 3 model, setting maximum lag length as 4 under SIC. As a result, the test finds ARDL(3,1,1,0,0,0,0,0,0,0) specification as most appropriate for BITP model where SIC value (10.8776) is the minimum (see Figure 5, Panel A). In other words, the SIC suggests that our ARDL model should include only three lags of dependent variables ($P_c$), one lag of *MARP* and *MARV*. For robustness of this ARDL model, we examine its residual under Serial Correlation LM and Heteroskedasticity test. We find that residuals do not comprise these two problems as probabilities of Chi-Square statistics of Breusch-Godfrey Serial Correlation LM test and Breusch-Pagan-Godfrey Heteroskedasticity test are greater than 11% and 12% respectively. However, the percentages are quite close to 10% significance level, therefore, we use HAC-robust standard errors in ARDL model. Further, we examine the stability of the model by employing CUSUM test that checks changes in cumulative sum of recursive residuals over time. In panel B of the figure 5, we demonstrate graphical outcome of CUSUM test where clearly seen that CUSUM (blue) line does not exceed ±5% significance (two red) lines, indicating stability of our model over the analysis period.

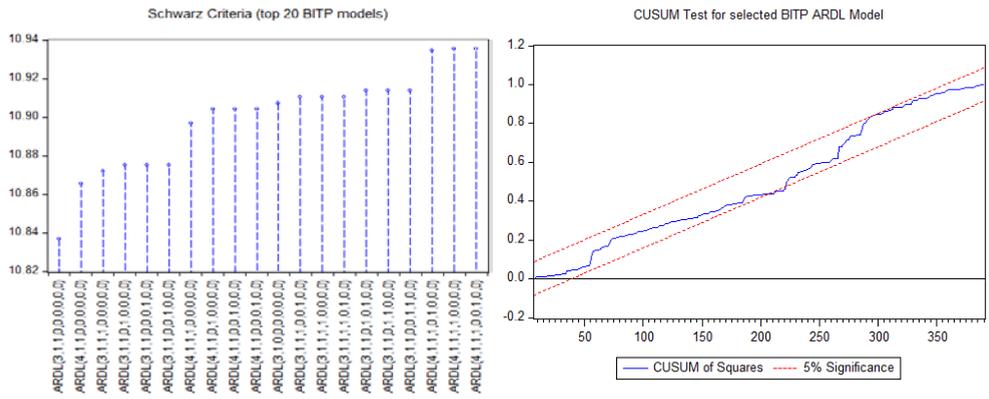

**Panel A.** ARDL Lag Selection with SIC          **Panel B.** CUSUM Test Result for ARDL Model

**Figure 5.** ARDL Lag Selection for BITP model with SIC & Stability of Selected ARDL Model

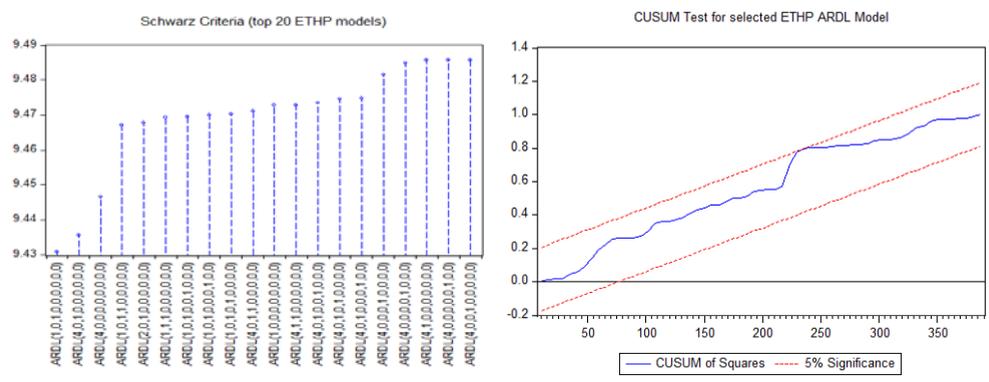

**Panel A.** ARDL Lag Selection with SIC          **Panel B.** CUSUM Test Result for ARDL Model

**Figure 6.** ARDL Lag Selection for ETHP model with SIC & Stability of Selected ARDL Model

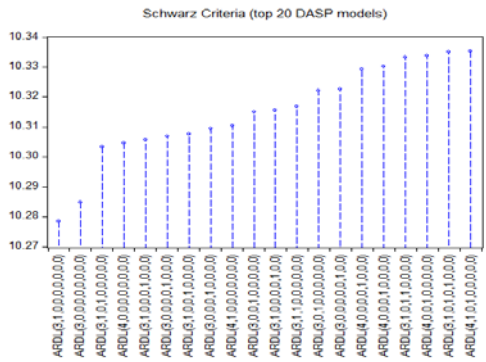
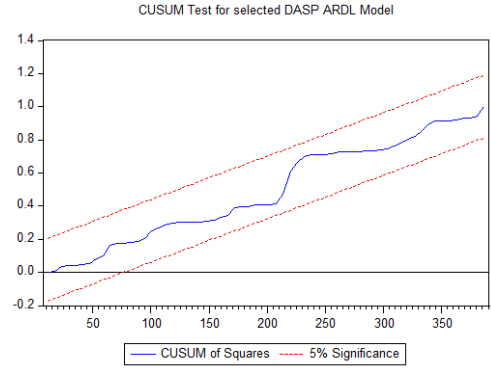

**Panel A.** ARDL Lag Selection with SIC    **Panel B.** CUSUM Test Result for ARDL Model
**Figure 7.** ARDL Lag Selection for DASP model with SIC & Stability of Selected ARDL Model

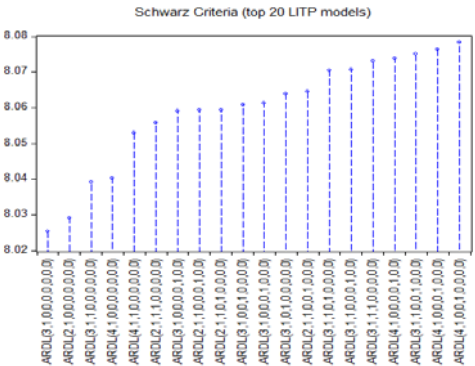
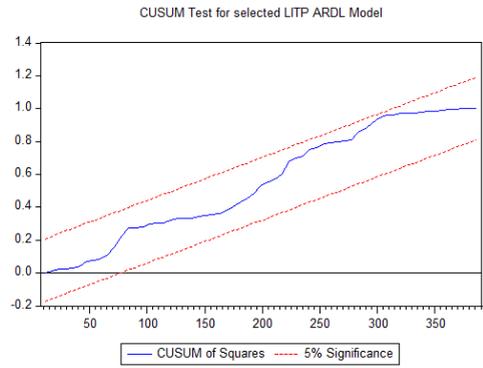

**Panel A.** ARDL Lag Selection with SIC    **Panel B.** CUSUM Test Result for ARDL Model
**Figure 8.** ARDL Lag Selection for LITP model with SIC & Stability of Selected ARDL Model

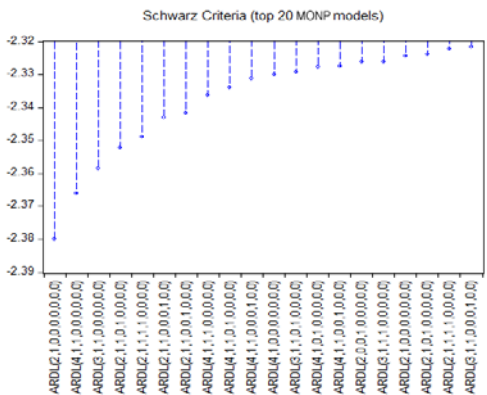
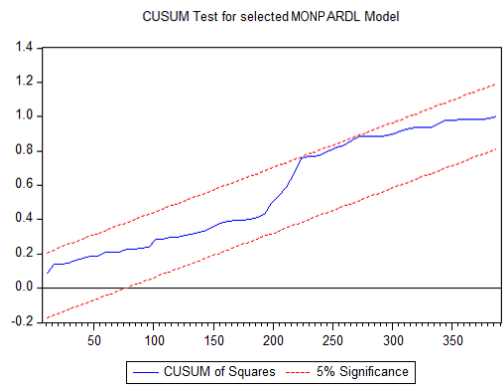

**Panel A.** ARDL Lag Selection with SIC    **Panel B.** CUSUM Test Result for ARDL Model
**Figure 9.** ARDL Lag Selection for MONP model with SIC & Stability of Selected ARDL Model

Similarly, on basis of minimum SIC value, we select the most appropriate ARDL models for *ETHP* (see figure 6), *DASP* (see figure 7), *LITP* (see figure 8), and *MONP* (see figure 9).

After specifying most appropriate models for our five cryptocurrency, we run equation 1 for each models separately. Table 4 reports outcome of these analyses where we observe plausible results. These models comprise both short- and long-run dynamics. In all cases $P_{t-1}$ derives statistically significant estimates at 1% level. We can derive long-run multipliers of related variables by using formula of $-\varphi_i/\varphi_1$ where $i$ has an array of {21, 22, 23, 24, 31, 32, 33, 34} that corresponds to {*MARP, MARV, MARS, ATR, EURP, GOLP, SPP, INT*}. However, we first need to carry out Wald test hypothesizing $H_0$: $\varphi_i=0$, to examine whether these series (cryptocurrency market variables and control variables) have statistically significant long-run interactions with $P_t$ (cryptocurrency *BITP, ETHP, DASP, LITP,* and *MONP*). We report results of Walt test in table 5 for each cryptocurrency where critical values for lower (*I(0)*) and upper (*I(1)*) bounds of each case also are given. Case I should be used for models that include neither intercept nor trends of any kind. Case II should be used for models that comprise only restricted intercept without any trends. In other words, the intercept is allowed only in long-run (Unrestricted ECM) model, but not short-run (Restricted ECM) model. On the other hand, case III allows unrestricted intercept, but no any trends. In this case, short-run model has an intercept and no trends. Case IV should be used for models that comprise both unrestricted intercept and restricted trend. So that short-run specification of these models includes intercept, while long-run specification comprises trend factor.

Therefore, specifying short- and long-run ARDL models is important. In this respect, we follow simple logic, we look if inclusion of intercept and trend add statistically significance into models or not. Once, we find they are statistically significant; we left them in the model. Following this methodology, we observe that trend factor appears insignificant in all cases (*BITP, ETHP, DASP, LITP,* and *MONP*). Thus, we disregard it in all cases. On the other hand, intercepts in *BITP*, *LITP*, and *MONP* models appear significant only in long-run, but not in short-run. Therefore, we choose case II specification for these three models. Intercept appears statistically insignificant only in *DASP* model, consequently removing it from the model we end up with case I specification.  Lastly, we observe that intercept in unrestricted in short-run model of *ETHP* as it derived statistically significance at 1% level. Thus, we choose case III specification for *ETHP* model.

**Table 4.** Results of ARDL Models

| Variable | ΔBITP | ΔETHP | ΔDASP | ΔLITP | ΔMONP |
|---|---|---|---|---|---|
| $\Delta P_{t-1}$ | 0.4208*** (0.0851) | - | 0.2362** (0.1165) | 0.2188*** (0.0799) | 0.9310*** (0.1865) |
| $\Delta P_{t-2}$ | 0.0260** (0.0131) | - | 0.3307** (0.1422) | 0.2297** (0.1187) | - |
| $\Delta MARP_t$ | 0.6966*** (0.0715) | 0.2111*** (0.0313) | 0.7837*** (0.0117) | 0.7573*** (0.0084) | 0.0484*** (0.0071) |
| $\Delta MARV_t$ | 0.1312*** (0.0328) | - | - | - | 0.0061*** (0.0015) |
| $MARP_{t-1}$ ($\phi_{21}$) | 0.0676*** (0.0209) | 0.0839*** (0.0320) | 0.0944** (0.0375) | 0.1419** (0.0618) | 0.1101** (0.0562) |
| $MARV_{t-1}$ ($\phi_{22}$) | 0.0121** (0.0055) | 0.0287** (0.0124) | 0.0107*** (0.0041) | 0.0265* (0.0154) | 0.0143 (0.0187) |
| $MARS_{t-1}$ ($\phi_{23}$) | -0.0128 (0.0139) | -0.0328 (0.0364) | -0.0205 (0.0214) | -0.0090 (0.0162) | -0.0028 (0.0032) |
| $ATR_{t-1}$ ($\phi_{24}$) | 0.1085*** (0.0311) | 0.0521*** (0.0162) | 0.0328 (0.0493) | 0.0288*** (0.0101) | 0.0205 (0.0171) |
| $EURP_{t-1}$ ($\phi_{31}$) | 0.0628 (0.0428) | 0.0376 (0.0909) | 0.0126 (0.0344) | 0.0164 (0.0380) | 0.0133 (0.0349) |
| $GOLP_{t-1}$ ($\phi_{32}$) | -0.0333 (0.0439) | 0.0131 (0.0374) | -0.0023 (0.0682) | 0.0109 (0.0597) | 0.0092 (0.0401) |
| $SPP_{t-1}$ ($\phi_{33}$) | 0.0696* (0.0368) | 0.0467* (0.0251) | 0.0349 (0.0316) | 0.0165* (0.0103) | 0.0122 (0.0169) |
| $INT_{t-1}$ ($\phi_{34}$) | -0.0317 (0.0535) | -0.0128 (0.0702) | -0.0049 (0.0172) | -0.0189 (0.0328) | -0.0001 (0.0007) |
| $P_{t-1}$ ($\phi_1$) | -0.0851*** (0.0299) | -0.2181*** (0.0529) | -0.8801*** (0.1179) | -0.4127*** (0.0780) | -0.4331*** (0.1091) |
| intercept | -0.0391*** (0.0118) | -0.0597*** (0.0276) | - | 0.0325*** (0.0035) | 0.0404*** (0.0060) |
| R-squared | 0.7122 | 0.5929 | 0.4703 | 0.5791 | 0.5454 |
| DurbinWatson | 2.0109 | 1.9282 | 2.0629 | 1.9929 | 1.9489 |
| BG LM Test | 0.1095 | 0.1413 | 0.0782 | 0.1276 | 0.0391 |
| BPG Test | 0.1216 | 0.0738 | 0.0655 | 0.0588 | 0.1302 |

**Notes:** Numbers in the table are estimations derived by ARDL technique with maximum 4 lags allowance. We use SIC in lag length selection. The standard errors are in HAC-robust characteristics with Bartlett kernel and Newey-West (1987) fixed bandwidth 5.

Table 5 presents Wald test of our five models comparatively with Pesaran et al (2001) critical values of lower and upper bounds for *k=8*. This *k* indicates number of original regressors in the model, except the dependent variable. The table shows that computed F-statistics exceeds critical bound values even at 1% significance level in all cases, signifying strong long-run cointegrations among mentioned series.

**Table 5.** F-test with Bound Critical Values

| **Bounds** | **10% level** | | **5% level** | | **1% level** | | **Wald Test** | |
|---|---|---|---|---|---|---|---|---|
| | Lower Bound [I(0)] | Upper Bound [I(1)] | Lower Bound [I(0)] | Upper Bound [I(1)] | Lower Bound [I(0)] | Upper Bound [I(1)] | | |
| **Case I** | 1.66 | 2.79 | 1.91 | 3.11 | 2.45 | 3.79 | DASP | (5.04***) |
| **Case II** | 1.85 | 2.85 | 2.11 | 3.15 | 2.62 | 3.77 | BITP | (8.96***) |
| | | | | | | | LITP | (6.40***) |
| | | | | | | | MONP | (5.28***) |
| **Case III** | 1.95 | 3.06 | 2.22 | 3.39 | 2.79 | 4.1 | ETHP | (7.53***) |
| **Case IV** | 2.13 | 3.09 | 2.38 | 3.41 | 2.93 | 4.06 | | |

**Notes:** The critical values for each case are retrieved from Pesaran et al (2001) Table CI. The $k$ indicates the number original regressors in the model. Therefore, it is 8 for in equation 1 model (disregarding dependent variable $P_{t-1}$). Wald test hypothesizes null $H_0: \varphi_i=0$. The significance levels respectively as *:10%, **:5%, and ***:1%.

Now, we can estimate approximate magnitude of cointegrations (long-run relationships) by calculating negative ratio of coefficients of independent variables to dependent one ($-\varphi_i/\varphi_1$). Table 6 shows these long-run multipliers automatically derived by Eviews software with their HAC-robust standard errors in parenthesis. According to results, we document that long-run market beta (coefficient of *MARP*) is statistically significant at 1% level in Bitcoin and Ethereum models where its multiplier is 0.79 and 0.38 respectively. Whereas, Dash, Litecoin, and Monero models predict it as 0.11, 0.34, and 0.25 respectively at 5% significance level. We believe that these results emerge as Bitcoin and Ethereum comprise the largest market share of entire cryptocurrency market, and their beta coefficient shows higher responsiveness to the market in long-run. In other words, 1 unit increase in *MARP* leads Bitcoin and Ethereum to increase by 0.79 and 0.38 units respectively.

On the other hand, trading volume appears to have significant long-run impact on Bitcoin at 1% significance level and on Ethereum, Litecoin, and Monero at 10% significance level. In case of Dash model, it appears statistically insignificant. The result indicates that a unit increase in weekly trading volume causes 0.14, 0.13, 0.06, and 0.03 increases in Bitcoin, Ethereum, Litecoin, and Monero cryptocurrencies in long-run. Sigma, proxied by volatility of the cryptocurrency market, emerges statistically significant long-run impact on all cryptocurrencies. The sign of impact is negative, which indicates a unit increase in volatility of the market causes Bitcoin to drop by 0.15 units, Ethereum by 0.15 units, Dash by 0.02, Litecoin by 0.02 units, and Monero by 0.01 units in long-run.

In addition, we observe that attractiveness (proxied by Google search term frequency) also derives significant coefficients for Bitcoin and Ethereum at 1% significance level and for Litecoin and Monero at 10% significance level. It indicates that 1 unit increase in Google trend popularity of Bitcoin, Ethereum, Litecoin, and Monero leads 1.27, 0.24, 0.07, and 0.05 units increases in their prices in long-run respectively. Google search frequency appears to be insignificant factor for Dash.

Coming to macroeconomic control variables, we observe that majority of them seem to be statistically insignificant factor in explaining price movements in cryptocurrencies. Only, SP500 index derives weak form of significant coefficient (10% level) in Bitcoin, Ethereum, and Litecoin models. The positive sign of SPP indicates that a unit increase leads 0.81, 0.21, 0.04 raise in Bitcoin, Ethereum, and Litecoin prices respectively in long-run. The logic behind this relationship appears ambiguous. Normally, one could expect a stronger USD against other fiat currencies (including cryptocurrencies) when SPP increases.

**Table 6.** Long-run estimates of Cryptocurrency ARDL models

|      | BITP        | ETHP        | DASP       | LITP       | MONP       |
|------|-------------|-------------|------------|------------|------------|
| MARP | 0.7944***   | 0.3847***   | 0.1073**   | 0.3438**   | 0.2542**   |
|      | (0.0599)    | (0.0654)    | (0.0519)   | (0.1649)   | (0.1202)   |
| MARV | 0.1425***   | 0.1316*     | 0.0122     | 0.0642*    | 0.0330*    |
|      | (0.0349)    | (0.0713)    | (0.0177)   | (0.0362)   | (0.0187)   |
| MARS | -0.1511***  | -0.1504***  | -0.0233*   | -0.0218**  | -0.0065**  |
|      | (0.0526)    | (0.0564)    | (0.0125)   | (0.0102)   | (0.0032)   |
| ATR  | 1.2750***   | 0.2389***   | 0.0373     | 0.0698**   | 0.0473**   |
|      | (0.1511)    | (0.0721)    | (0.0328)   | (0.0334)   | (0.0197)   |
| EURP | 0.7381***   | 0.1724      | 0.0143     | 0.0397     | 0.0307     |
|      | (0.2446)    | (0.1149)    | (0.0385)   | (0.0502)   | (0.0349)   |
| GOLP | -0.3913     | 0.0601      | -0.0026    | -0.0264    | -0.0212    |
|      | (0.2788)    | (0.0795)    | (0.0681)   | (0.0681)   | (0.0401)   |
| SPP  | 0.8179*     | 0.2141*     | 0.0397     | 0.0400*    | 0.0282     |
|      | (0.4250)    | (0.1133)    | (0.0324)   | (0.0213)   | (0.0184)   |
| INT  | -0.3725     | -0.0587     | -0.0056    | -0.0458    | -0.0002    |
|      | (0.2357)    | (0.0683)    | (0.0243)   | (0.0464)   | (0.0171)   |
| C    | 0.4596***   | -           | -          | 0.3668***  | 0.4284***  |
|      | (0.0827)    |             |            | (0.0739)   | (0.0817)   |

**Notes:** Estimates are derived by long-run unrestricted ARDL technique with HAC-robust standard errors in parenthesis. The significance levels are: 10% (*), 5% (**), and1% (***).

Further, we estimate short-run error-correction equation (4) and we report outcome of this analysis in table 7. Apparently, all ECM models generate consistent coefficients. The coefficient of *MKT* implies that a unit increase in cryptocurrency market return causes Bitcoin, Ethereum, Dash, Litecoin, and Monero to increase by 0.85, 0.39, 0.04, 0.12, and 0.09 units respectively in short-run. Notice that short-run coefficients of Bitcoin and Ethereum are higher comparing to their long-run coefficients, indicating that their responses are more sensitive in the short-run. Likewise, a unit increase in cryptcurrency market trading volume leads mentioned cryptocurrencies to increase by 0.03, 0.01, 0.007, 0.005, and 0.004 units respectively at 1%-5% significance level. These short-run coefficients seem to be lesser than their long-run magnitudes at table 6, indicating that responses of the cryptocurrencies to the fluctuations in market trading volume are higher in long-run.

**Table 7.** Short-run estimates of ARDL Error-Correction Cryptocurrency models

|  | BITP | ETHP | DASP | LITP | MONP |
|---|---|---|---|---|---|
| $\Delta P_{t-1}$ | 0.2087*** (0.0429) | - | 0.1585*** (0.0545) | 0.3973*** (0.0651) | 0.5271*** (0.1163) |
| $\Delta P_{t-2}$ | 0.1251*** (0.0115) | - | 0.1162*** (0.0322) | 0.1862*** (0.0642) | - |
| $\Delta$MARP | 0.8485*** (0.0983) | 0.3914*** (0.0457) | 0.0414** (0.0187) | 0.1195** (0.0531) | 0.0874** (0.0428) |
| $\Delta$MARV | 0.0315*** (0.0022) | 0.0113*** (0.0011) | 0.0075** (0.0034) | 0.0053** (0.0027) | 0.0041** (0.0020) |
| $\Delta$MARS | -0.3896** (0.1902) | -0.2409* (0.1453) | -0.1695* (0.0913) | -0.1991** (0.0977) | -0.1347* (0.0735) |
| $\Delta$ATR | 0.1372* (0.0817) | 0.0538 (0.0644) | 0.0270 (0.0451) | 0.0257 (0.0211) | 0.0288 (0.0436) |
| $\Delta$EURP | 0.0723 (0.1131) | 0.0473 (0.0899) | 0.0148 (0.0595) | 0.0245 (0.0395) | 0.0280 (0.0409) |
| $\Delta$GOLP | 0.1399 (0.1664) | -0.0023 (0.0962) | 0.0768 (0.1452) | 0.0462 (0.0565) | 0.0301 (0.0403) |
| $\Delta$SPP | -0.2020* (0.1328) | -0.0683 (0.0676) | 0.0410 (0.0718) | -0.0360 (0.0328) | -0.0203 (0.0302) |
| $\Delta$INT | -0.0597 (0.1106) | -0.0234 (0.0844) | -0.0022 (0.0087) | -0.0175 (0.0314) | -0.0019 (0.0058) |
| C | - | 1.1541*** (0.2118) | - | - | - |
| $ECT_{t-1}$ | -0.2368*** (0.0361) | -0.1276*** (0.0303) | -0.1020*** (0.0212) | -0.2291*** (0.0472) | -0.1427*** (0.0335) |

**Notes:** The estimates are derived by short-run RECM ARDL model with HAC-robust standard errors in parenthesis. *ECT_{t-1}* shows speed of adjustment towards long-run equilibrium. The significance levels are: 10% (*), 5% (**), and 1% (***). For data description and abbreviations see table 2.

Market volatility also appears statistically significant and negatively signed as it was in long-run. Moreover, short-run magnitudes of all cryptocurrencies are several-fold comparing to their long-run magnitudes, signalizing that the cryptocurrencies show more severe reaction to the market's volatility in short-run.

Interestingly, attractiveness factor derives insignificant estimates for almost all models, except Bitcoin that predicts 0.14 coefficient at 10% significance level. This indicates that impact of attractiveness on cryptocurrency is subjected to time factor, indicating that it has latent characteristics and its formation (may be also recognition by the market) requires a time.

On the other hand, macroeconomic control variables appear insignificant in all short-run models, except Bitcoin model that predicts estimate of *SPP* factor as -0.2020 at 10% significance level. This indicates that a unit increase in SP500 index causes Bitcoin prices to decrease by 0.20 units in short-run. An inverse relationship is documented in long-run with absolute magnitude of several-fold. This, indeed, confirms negative correlation between Bitcoin prices and SP500 index in short-run.

Lastly, error correction terms (*ECT*) in all models appear statistically significant at 1% level with negative sign complying with the ECM theory. Bitcoin model seems to be correcting 23.68% of its previous period disequilibrium in the way converging its long-run level. This adjustment speed in Etherem, Dash, Litecoin, and Monero is 12.76%, 10.20%, 22.91%, and 14.27% respectively.

## 6. Concluding remarks

This paper examines factors that influence prices of most common five cryptocurrencies such Bitcoin, Ethereum, Dash, Litecoin, and Monero over 2010-2018 using weekly data and documents several results. First, using differencing methodology to stationarize series wipes out potential long-run interactions between series; therefore, we use Autoregressive Distributed Lag (ARDL) technique in order to account both short- and long-run dynamics of cryptocurrency prices as our data sample is comprised of mixture of I(0) and I(1) variables. Unrestricted long-run ARDL and restricted short-run error-correction analyses find statistically significant impact running from cryptomarket factors such as total market prices, trading volume, and volatility on to five cryptocurrencies in long- and short-run respectively.

The cryptomarket beta derives a long-run multiplier of 0.79 on Bitcoin and 0.38 on Ethereum at 1% significance level, while it generates 0.11, 0.34, and 0.25 long-run impacts on Dash, Litecoin, and Monero at 5% significance level. This indicates that Bitcoin and Ethereum have higher responsiveness to the market in long-run. In case of short-run, a unit increase in cryptocurrency market return causes Bitcoin, Ethereum, Dash, Litecoin, and Monero to increase by 0.85, 0.39, 0.04, 0.12, and 0.09 units respectively in short-run. As short-run multiplier of Bitcoin and Ethereum are greater than their long-run coefficients, we conclude that these responses of these two cryptocurrencies are more sensitive in short-run.

Trading volume appears to have significant long-run impact on Bitcoin at 1% significance level and on Ethereum, Litecoin, and Monero at 10% significance level, indicating a unit increase in weekly trading volume causes 0.14, 0.13, 0.06, and 0.03 raises in Bitcoin, Ethereum, Litecoin, and Monero cryptocurrencies in long-run. In case of short-run dynamics, all five cryptocurrencies earn statistically significant estimates. However, these estimates seem to be lesser than their long-run magnitudes, indicating that responses of the cryptocurrencies to the fluctuations in market trading volume are higher in long-run.

Likewise, volatility of the cryptocurrency market appears to be statistically significant determinant both in long- and short-runs for all cryptocurrencies. The sign of impact is negative, which indicates a unit increase in volatility of the market causes Bitcoin to drop by 0.15 units, Ethereum by 0.15 units, Dash by 0.02, Litecoin by 0.02 units, and Monero by 0.01 units in long-run. In case of short-run, these impacts seem to be several-fold, indicating that the cryptocurrencies show more severe reaction to the market's volatility in short-run.

Attractiveness of cryptocurrencies also matters for all except Dash, but only in long-run. It derives significant coefficients for Bitcoin and Ethereum at 1% significance level and for Litecoin and Monero at 10% significance level, indicating that 1 unit increase in attractiveness of Bitcoin, Ethereum, Litecoin, and Monero leads 1.27, 0.24, 0.07, and 0.05 units increases in their long-run prices respectively. In case of short-run analysis, attractiveness factor derives insignificant estimates for almost all models, except Bitcoin that earns an estimate of 0.14 at 10% significance level. This indicates that formation and recognition of the attractiveness of cryptocurrencies are subjected to time factor. In other words, they travel slowly within the market.

In case of control variables, SP500 index derives weak form of positive significant coefficient (10% level) in Bitcoin, Ethereum, and Litecoin models. Although the logic behind these positive long-run relationships appears ambiguous, they totally disappear in short-run, while only Bitcoin model predicts a negative estimate that is statistically significant 10% significance. This confirms that one could expect a stronger USD against other fiat currencies (including cryptocurrencies) when SPP increases.

Lastly, error correction terms (ECT) in all models appear statistically significant at 1% level with negative sign complying with the ECM theory. Bitcoin model seems to be correcting 23.68% of its previous period disequilibrium in the way converging its long-run level. This adjustment speed in Etherem, Dash, Litecoin, and Monero is 12.76%, 10.20%, 22.91%, and 14.27% respectively.

The main limitation of the study is latency (novelty, obscurity, and intangibilty) of majority of cryptocurrency related information. This is a brand new market and a brand new topic for academic researches. We also believe that if we could proxy adaptation of cryptocurrency (legalization of any cryptocurrency as a payment tool), we believe it would improve our model further. In addition, few cryptocurrencies comprise speculative bubbles, particularly Bitcoin, thus, future researches should attempt to measure volume of this bubble addressing to question *"are we in the peak of Bitcoin bubble?"*.

**APPENDIX**

## Table 1A. Bitcoin Legality by Country and Classification

| | Name | Bitcoin Legality | Classification | | Name | Bitcoin Legality | Classification |
|---|---|---|---|---|---|---|---|
| 1 | Afghanistan | Illegal | Currency | 59 | Lebanon | Legal | No Information |
| 2 | Aland Islands | Legal | Currency | 60 | Liberland | Legal | Currency |
| 3 | Algeria | Illegal | Currency | 61 | Libyan Arab Jamahiriya | Legal | Money |
| 4 | American Samoa | Restricted | Commodity | 62 | Liechtenstein | Legal | Currency |
| 5 | Andorra | Neutral / Alegal | No Information | 63 | Lithuania | Legal | Currency |
| 6 | Argentina | Neutral / Alegal | Property | 64 | Luxembourg | Legal | Currency |
| 7 | Australia | Legal | Currency | 65 | Malaysia | Neutral / Alegal | No Classification |
| 8 | Austria | Legal | Currency | 66 | Maldives | Neutral / Alegal | No Information |
| 9 | Azerbaijan | Legal | Currency | 67 | Malta | Legal | Currency |
| 10 | Bangladesh | Illegal | No Information | 68 | Mauritius | Neutral / Alegal | No Classification |
| 11 | Barbados | Neutral / Alegal | No Information | 69 | Mexico | Restricted | Currency |
| 12 | Belarus | Legal | No Information | 70 | Monaco | Legal | Currency |
| 13 | Belgium | Legal | Currency | 71 | Mongolia | Legal | No Information |
| 14 | Bolivia | Illegal | No Information | 72 | Morocco | Illegal | No Information |
| 15 | Brazil | Legal | Commodity | 73 | Nepal | Restricted | No Classification |
| 16 | Brunei Darussalam | Legal | Currency | 74 | Netherlands | Legal | Commodity |
| 17 | Bulgaria | Legal | Currency | 75 | New Zealand | Legal | Commodity |
| 18 | Canada | Legal | Barter Good | 76 | Nicaragua | Legal | No Information |
| 19 | Chile | Legal | No Information | 77 | Nigeria | Neutral / Alegal | Currency |
| 20 | China | Restricted | Commodity | 78 | Northern Mariana Islands | Legal | Commodity |
| 21 | Colombia | Neutral / Alegal | No Classification | 79 | Norway | Legal | Commodity |
| 22 | Congo | Legal | No Information | 80 | Pakistan | Neutral / Alegal | No Classification |
| 23 | Costa Rica | Legal | Currency | 81 | Paraguay | Neutral / Alegal | No Classification |
| 24 | Croatia | Legal | Currency | 82 | Peru | Neutral / Alegal | No Classification |
| 25 | Cuba | Legal | Currency | 83 | Philippines | Legal | Barter Good |
| 26 | Cyprus | Legal | Currency | 84 | Poland | Legal | Property |
| 27 | Czech Republic | Legal | Currency | 85 | Portugal | Legal | No Classification |
| 28 | Denmark | Legal | Currency | 86 | Republic of Macedonia | Illegal | No Information |
| 29 | Ecuador | Illegal | No Information | 87 | Reunion | Legal | Commodity |
| 30 | Egypt | Restricted | Commodity | 88 | Romania | Legal | Currency |
| 31 | Estonia | Legal | Currency | 89 | Russian Federation | Illegal | Currency |
| 32 | Finland | Legal | Currency | 90 | San Marino | Legal | Currency |
| 33 | France | Legal | Commodity | 91 | Saudi Arabia | Restricted | No Information |
| 34 | Gabon | Neutral / Alegal | No Information | 92 | Serbia | Legal | No Information |
| 35 | Georgia | Legal | No Classification | 93 | Singapore | Legal | Currency |
| 36 | Germany | Legal | Barter Good | 94 | Slovakia | Legal | Currency |
| 37 | Greece | Legal | Currency | 95 | Slovenia | Legal | Currency |
| 38 | Hong Kong | Legal | Commodity | 96 | South Africa | Legal | Currency |
| 39 | Hungary | Legal | Currency | 97 | South Korea | Legal | No Classification |
| 40 | Iceland | Legal | Currency | 98 | Spain | Legal | Currency |
| 41 | India | Neutral / Alegal | Commodity | 99 | Svalbard and Jan Mayen | Legal | Commodity |
| 42 | Indonesia | Neutral / Alegal | Commodity | 100 | Sweden | Legal | Commodity |
| 43 | Iran | Legal | No Classification | 101 | Switzerland | Legal | Currency |
| 44 | Iraq | Legal | No Information | 102 | Taiwan | Legal | No Information |
| 45 | Ireland | Legal | Currency | 103 | Thailand | Legal | Commodity |
| 46 | Isle of Man | Legal | No Information | 104 | Tunisia | Neutral / Alegal | No Classification |
| 47 | Israel | Legal | Commodity | 105 | Turkey | Legal | Commodity |
| 48 | Italy | Legal | Currency | 106 | Ukraine | Legal | Currency |
| 49 | Japan | Legal | Currency | 107 | United Arab Emirates | Legal | Currency |
| 50 | Jersey | Legal | Currency | 108 | United Kingdom | Legal | Currency |
| 51 | Jordan | Neutral / Alegal | No Classification | 109 | United States of America | Legal | Property |
| 52 | Kazakhstan | Neutral / Alegal | Currency | 110 | Uruguay | Neutral / Alegal | Property |
| 53 | Kenya | Neutral / Alegal | No Classification | 111 | Uzbekistan | Legal | Currency |
| 54 | Kosovo | Neutral / Alegal | No Information | 112 | Venezuela | Neutral / Alegal | Commodity |
| 55 | Kuwait | Legal | No Information | 113 | Viet Nam | Neutral / Alegal | Property |
| 56 | Kyrgyzstan | Neutral / Alegal | Currency | 114 | Zambia | Restricted | No Information |
| 57 | Latvia | Legal | Currency | 115 | Zimbabwe | Legal | Commodity |